\documentclass[letterpaper,twocolumn,superscriptaddress,showkeys]{revtex4}
\usepackage[latin1]{inputenc}
\usepackage{amsmath,dcolumn,graphicx,hyperref}

\hypersetup {
  colorlinks = true, linkcolor = blue, citecolor = blue, urlcolor = blue, pdfauthor = {Desjardins-Proulx, Philippe},
}

\begin{document}

\title{How likely is speciation in neutral ecology ?}

\author{Philippe Desjardins-Proulx}
\email[E-mail: ]{philippe.d.proulx@gmail.com}
\affiliation{College of Engineering, University of Illinois at Chicago, USA.}
\affiliation{Canada Research Chair on Terrestrial Ecosystems, Universit\'e du Qu\'ebec, Canada.}

\author{Dominique Gravel}
\affiliation{Canada Research Chair on Terrestrial Ecosystems, Universit\'e du Qu\'ebec, Canada.}

\begin{abstract}
Patterns of biodiversity predicted by the neutral theory rely on a simple phenomenological model of speciation. To further investigate the effect of speciation on neutral biodiversity, we analyze a spatially-explicit neutral model based on population genetics. We define the metacommunity as a system of populations exchanging migrants and we use this framework to introduce speciation with little or no gene flow (allopatric and parapatric speciation). We find that with realistic mutation rates, our metacommunity model driven by neutral processes cannot support more than a few species. Adding natural selection in the population genetics of speciation increases the number of species in the metacommunity but the level of diversity found in Barro Colorado Island is difficult to reach.
\end{abstract}


\maketitle

\section{Introduction}

How patterns of biodiversity arise through ecological and evolutionary processes is a central question in modern ecology \cite{joh07,fus07}. According to Hubbell's neutral theory of biodiversity (NTB), patterns of biodiversity such as species-abundance distributions can be explained by the balance between speciation, dispersal and random extinction \cite{hub01,ros11b}. The neutral theory provides a good fit to species distribution curves \cite{hub01} and has been extended in several ways \cite{hae09,vol05,dea09,ros10}. The neutral theory is flexible enough to fit nearly any species abundance distribution \cite{cha02}, but, species abundance distributions apart, it provides valid starting points and interesting null hypotheses for many problems in community ecology \cite{alo06,lei07}.

While a lot has been said about the assumption of ecological equivalence \cite{abr01b,pur10}, much less attention has been given to the speciation mode \cite{eti07}, which is sometime seen as the theory's weakest point \cite{kop10}. In recent years, several variants of the NTB have explored different speciation models \cite{eti07,hae09,ros10,dea09}. However, nothing has been done to relate the theory to population genetics and known models of speciation, despite the fact that, as Etienne et al. noted \cite{eti07}, such a mechanistic model could eventually force us to reject neutrality. The neutral theory with point speciation has also been criticized for predicting too many rare species, too many young species \cite{ric03}, and for assuming a direct relationship between abundance and speciation \cite{eti07}. 

In this article, we introduce a neutral theory of biodiversity with a speciation model derived from population genetics. We emphasize the role of allopatric and parapatric speciation. Speciation modes are most often distinguished according to the level of gene flow between the diverging populations. Allopatric speciation occurs when the new species originates from a geographically isolated population. By contrast, sympatric speciation is often defined as speciation without geographical isolation, in short, when the diverging populations share the same location. Lastly, parapatric speciation covers the middle ground between these two extremes \cite{gav03}.

In the original neutral theory's formulation, Hubbell presented two models of speciation, point speciation and random fission speciation \cite{hub01}. Both are phenomenological individual-based models. In the case of point speciation, a newly recruited individual is selected at random and undergoes speciation. In the case of random fission, the whole species is divided in two at random. The random fission model is more realistic and does improve some predictions related to speciation, but the resulting species abundance curves do not fit data as well as the point speciation model \cite{eti11}. In both cases, the probability of speciation of a given species is directly proportional to abundance and independent of dispersal. Hubbell associates the point speciation model with sympatric speciation, and the random fission model with allopatric speciation \cite{hub01}. Some rare forms of sympatric speciation are indeed similar to the point speciation model, namely polyploid speciation, but most sympatric speciation events involve a population being divided in two by non-geographical factors \cite{coy04}. Also, as neither model takes gene flow into consideration, neither can distinguish sympatric and allopatric speciation events.

While theoretical models have shown sympatric speciation to be possible, empirical studies have uncovered very few solid cases \cite{bol07} and much of the theory is controversial \cite{spe05,bar05}. Despite the growing acceptance of sympatric speciation as a plausible cause of speciation, most speciation events are still thought to occur with limited gene flow \cite{coy04,gav05,bol07,fit08}. Sympatric speciation is difficult to achieve for two reasons \cite[p. 127]{coy04}. First, the antagonism between selection and recombination. As selection pushes the populations in different directions, gene flow tends to break combinations that would be beneficial for one population but not the other, creating maladapted genotypes. Also, the diverging populations have to coexist before and after reproductive isolation. Allopatric and parapatric speciation events are thought to be more common but modelling them require some details about the spatial structure of the metacommunity. Ricklefs argued that allopatric speciation is the creative force in community ecology \cite{ric08} and we choose to base our model on the most common forms of speciation despite the increased complexity of a spatially-explicit framework. We find that with realistic parameters, metacommunities cannot support more than a few species when the genetics of speciation is assumed to be neutral. We also considered a simple alternative pseudo-selection model by adding natural selection at the genetic level, but keeping the ecological equivalence assumption at the individual level. This model shows that the rates of speciation typical of the NTB cannot be obtained without selection pushing mutations to fixation.

\section{Model}

\begin{figure}[ht!]
\centering
\includegraphics[width=0.45\textwidth]{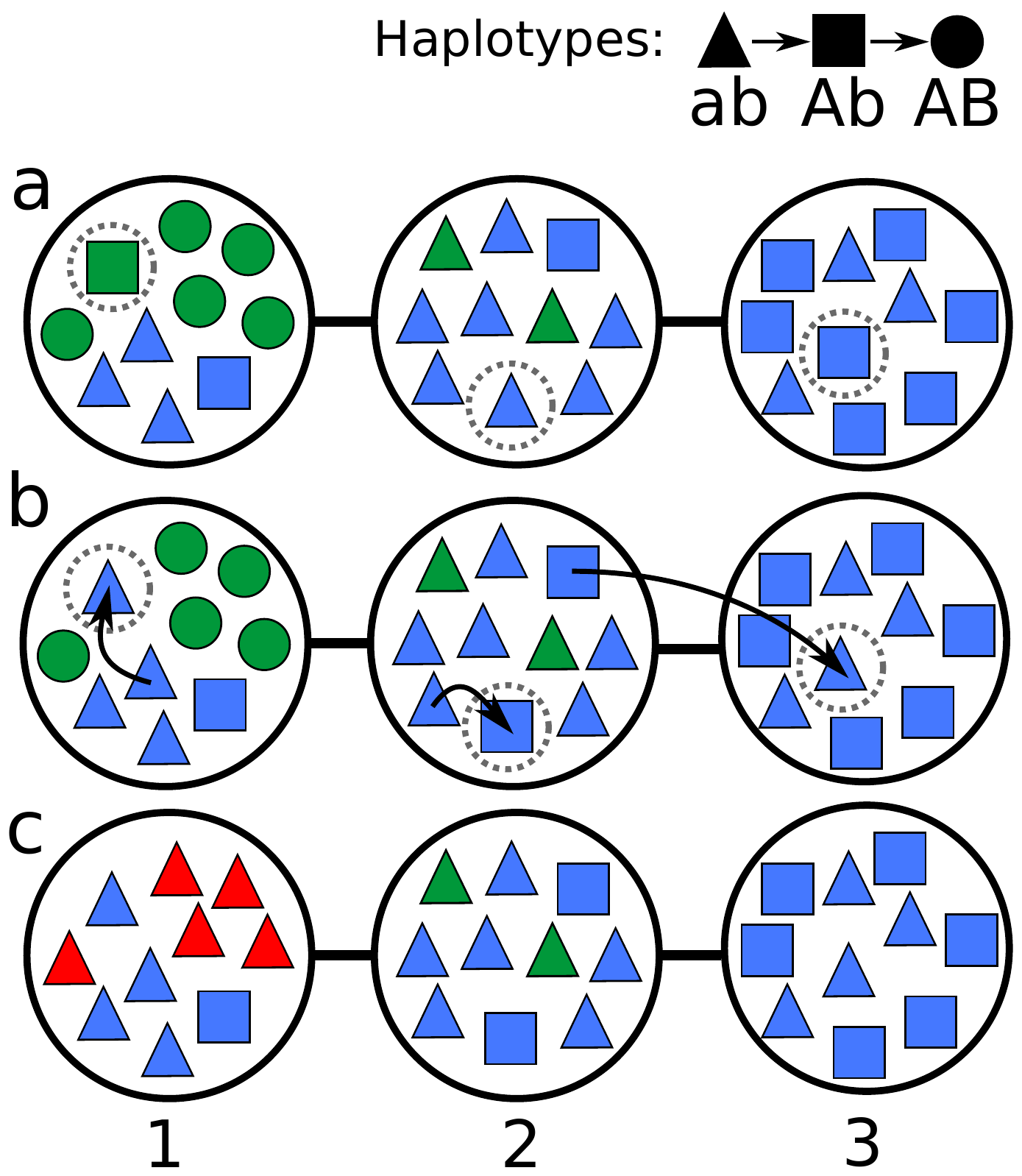}
\caption{
Illustration of the dynamics of a metacommunity with three local communities (numbered from 1 to 3) of 10 individuals. Colors and shapes are used to distinguish species and haplotypes, respectively. (a) At each time step, an individual is selected in each community. All individuals have the same probability $1/J_x$ of being selected, with $J_x$ being the size of the local community. (b) The individuals selected are then replaced either by migration (with probability $m$) or by local replacement (with probability $1-m$). In community 1, the individual is replaced by a local replacement event. A blue individual is chosen with $P(blue) = 0.4$, and then the $a_1b_1$ haplotype is chosen with $P(a_1b_1) = 3w_{a_1b_1}/(3w_{a_1b_1} + 1w_{A_1b_1})$, with $w$ being the fitness of the various haplotypes. In community 2, a blue individual with haplotype $a_2b_2$ is selected and mutates to $A_2b_2$ (probability $\mu$). The individual in community 3 is replaced by a migrant. A blue individual is selected in community 2 with probability $0.8$. While this individual carries $A_2b_2$, we assume different mutations are required in each community to achieve speciation. Thus, migrants carry no mutations at the focal loci for the population into which they move and the $A_2b_2$ haplotype is irrelevant in community 3. (c) At the end of the time step, speciation occurs if $A_xB_x$ is fixed. Because all green individuals in the community 1 carry $A_1B_1$, they speciate and are now represented by red triangles ($a_1b_1$). 
}
\end{figure}

We model speciation with the Bateson-Dob\-zhan\-sky-Muller model (BDM) in which reproductive isolation is caused by the accumulation of incompatible alleles \cite{bat09,dob37,mul42,orr96b,orr01}. While the BDM model is simple, we have many empirical and theoretical reasons to think that speciation events often follow a similar scheme \cite{gav03,coy04}. We use a two-loci and two-alleles version of the model where sexual reproduction is ignored \cite[p. 131]{gav04}. A species can be divided into several populations living in different communities, with each population having its own set of incompatible alleles at different loci. A population in community $x$ starts with the $a_xb_x$ haplotype fixed. The allele at the first locus, $a_x$, mutates to $A_x$, and the allele at the second locus, $b_x$, mutates to $B_x$. Both mutates at the same rate $\mu$. We follow Gavrilets and ignore back mutations \cite{gav04}. Back mutations have been shown to slow down speciation in this model, but not dramatically \cite[p. 131]{gav04}. The path from $a_xb_x$ to $A_xB_x$ can be seen as a process with three states: 

\begin{equation}\label{state}
	a_xb_x \longrightarrow A_xb_x \longrightarrow A_xB_x.
\end{equation}

$a_xB_x$ is absent because of an incompatibility between $a_x$ and $B_x$. Speciation occurs when all individuals in the population carry the $A_xB_x$ haplotype. Migration brings new individuals, always with the $a_xb_x$ haplotype, at rate $m$. To integrate Gavrilet's BDM model in a metacommunity, we connect local communities composed of populations of one or more species. Speciation is a complex process, but this simple model captures many important characteristics of speciation events that are ignored in the NTB. First, speciation takes time. It is the result of a long process where a population diverges from the rest of the species to the point where reproductive isolation prevents them from producing fertile progenies \cite{coy04}. Second, with a few exceptions, the starting population size of the new species is likely to be higher than one \cite{gav04,ros10}. Third, gene flow (migration) has a strong homogenizing effect that will inhibit speciation \cite{coy04,fit08}. Lastly, speciation occurs as a population of a given species diverges, most often in well-defined geographic areas \cite{avi00,coy04}. None of these characteristics are present in the original neutral theory \cite{hub01}, although protracted speciation partially solves the first two problems by adding a parameter to account for the duration of speciation \cite{ros10}. The first two problems were also solved within a different framework with assortative mating \cite{dea09,mel10}.

It is difficult to distinguish populations in individual-based models. As speciation is the result of divergences between populations, it is hard to model unless individuals are grouped into populations \cite{coy04}. In the NTB and most of its variants, only two levels of organization are recognized; the individual and the species. To integrate Gavrilets' model in the NTB, we model populations in patches using graphs. Several approaches have been used to model the spatial structure of populations and local communities. Some are spatially-explicit at the level of the individual. In these models, the location of each individual is known, generally by using a grid \cite{ros07} or a graph \cite{lie05}. Another approach is to consider the position of populations, but ignore the exact position of the individuals within the populations. Again, this method has been used with grids \cite{gav05} and graphs \cite{min07,eco08,dal10}. We use the latter approach and model the metacommunity as a graph of $n$ local communities (hereafter simply referred to as communities), where each community $x$ can support a total of $J_x$ individuals. These communities, composed of one or more species, are connected by dispersal \cite{eco08,eco10} (Fig. 1). This spatial representation allows us to distinguish three levels of organization: species, populations, and individuals. A population is simply the sum of all the individuals of a given species in a given community. A species can thus be divided in up to $n$ populations. Dispersal between two communities will always be low enough to assume that the individuals in these two communities can be defined as distinct populations \cite{ber02}.

All individuals have a haplotype (either $a_xb_x$, $A_xb_x$, $A_xB_x$) and we follow explicitly their dynamics in each community. As these haplotypes represent a path toward speciation in a particular community, they should be seen as different pairs of loci for each population. For example, if an individual migrates from community $1$ to community $2$, it will carry the $a_2b_2$ haplotype in its new community, regardless of its haplotype in community $1$. This assumption is not realistic in all situations, as both mutational-order and ecological speciation are known to be influenced by complex interactions between the diverging populations \cite{man90,sch09,nos11,coy04,gav04}. Integrating the effect of these divergences would require many more assumptions about the nature of speciation, and in most cases cannot be done without introducing the concept of niche \cite{sch00}. We ignore much of the details of speciation in favor of a simple model that captures many of the most fundamental characteristics of speciation as a population-process \cite{gav04}. Because there is no niche differentiation, new mutations toward speciation are always allowed to appear regardless of the ecological context. As soon as all the individuals of a given species inside a local community  $x$ carry the haplotype $A_xB_x$, they undergo speciation (Fig. 1). An alternative approach would be to allow $A_xB_x$ individuals to underdo speciation even in the presence of $a_xb_x$ if there are no $A_ib_i$ present. However, this model would fail to account for the homogenizing effect of gene flow and would almost always lead to sympatric speciation. As we want to model allopatric and parapatric speciation, we follow Gavrilets \cite{gav04} and only allow speciation when $A_xB_x$ is fixed.

Metacommunity dynamics is similar to Hubbell's neutral model of biodiversity \cite{hub01} and the Moran model in population genetics \cite{mor62,ewe04}. It can be described in three steps (Fig. 1). (1) For each time step, an individual is selected in each community. All individuals have the same probability $1/J_x$ of being selected, with $J_x$ being the size of the community (Fig. 1a). (2) The individuals selected in step 1 are replaced either by migration or by local replacement (Fig. 1b). The probability of migration from $x$ to $y$ is given by the matrix $\mathbf{m}$. In the case of migration from $x$ to $y$ ($x \not = y$), the new individual will belong to species $i$ with probability $N_{ix}/J_x$, with $N_{ix}$ being the population size of species $i$ in community $x$. We assume that migrants carry no mutations at the focal loci for the population into which they move so the haplotype is ignored and the new individual will carry $a_yb_y$. In the case of local replacement events, the new individual will also belong to species $i$ with probability $N_{ix}/J_x$. However, the fitness of the haplotypes is used to determine the new individual's haplotype. In the neutral model we can simply select the species and haplotype using relative abundance. One of the basic tenets of the NTB is ecological equivalence, so to introduce selection within the framework of neutral ecology the probability to pick an individual from one species has to ignore the internal genetic composition. In this pseudo-selection model, we use a multiplicative fitness regime \cite[p. 166]{cha11}, leading to $w_{a_xb_x} = 1$, $w_{A_xb_x} = 1 + s$, and $w_{A_xB_x} = (1+s)^2$, where $w$ denotes fitness and $s$ is the selection coefficient. The neutral model is the special case $s = 0$. In reality, if a population has many individuals with haplotypes $A_xb_x$ and $A_xB_x$, it should have an advantage over a population with only $a_xb_x$ individuals, but this would break the ecological equivalence assumption of neutral ecology so we ignore it. After the haplotype is selected, $a_xb_x$ mutates to $A_xb_x$ and $A_xb_x$ to $A_xB_x$ at rate $\mu$. (3) In the last step, all populations with $A_xB_x$ fixed undergoe speciation (Fig. 1c). The individuals of the new species will carry $a_xb_x$ and a new path toward speciation is possible. This is similar to the infinite sites approach of population genetics \cite{cro70} and is a direct consequence of neutrality.

We consider four different metacommunity shapes; circle, complete, star, and random. In the circle each community is linked by migration to its two neighbouring communities. In the complete metacommunity, each community is linked to all the others. In the star, a single central community forms a link to all outer communities, which have no other links. The random graph is an assemblage of communities based on random geometric graphs \cite{pen03}. These random graphs are used to test algorithms designed for spatial structures such as maps \cite{sed02}. The migration matrix is built with a single parameter $\omega$, which represent the strength of the links between communities. The migration probability between two linked communities $x$ and $y$ is found by dividing $\omega$ by the sum of all links to community $x$ plus one (for local replacement events). This method ensures that all rows in the migration matrix sum to 1 and that communities with more links are subjected to stronger migration. The probability that an individual selected in community $x$ will be replaced by migration is

\begin{equation}
  m_x = \frac{c\omega}{1 + c\omega} \approx c\omega,
\end{equation}

with $c$ being the number of communities linked to $x$. The 1 in the denominator stands for the weight given to local replacement events. $\omega$ is always much smaller than 1 so the average migration probability is approximately $c\omega$. Circle communities all have $c = 2$, for communities in the complete metacommunity it is $c = n - 1$, and for stars we have $c = 1$ except for the central community where $c = n - 1$. The average number of links for the random graphs depends on $n$ but vary little for $n < 30$. With $n = 10$, the random graphs have on average $2.56$ links.

We explored the model by simulations using an implementation in ANSI C99. Each simulation starts with 20 species evenly distributed in the metacommunity. We compared simulations with communities of size $J_x = 10^2$ to $10^6$ and found similar results. We thus use $J_x = 10^4$ unless otherwise noted. The mutation rate $\mu$ for eukaryotes is generally between $10^{-4}$ and $10^{-6}$ \cite{dra98,gav04} and we set $\mu$ to the highest realistic value, $\mu = 10^{-4}$. The simulations ran for 100 000 generations (a generation being $J_x$ time steps \cite{ros10}). We recorded the average local and regional species richness over the last 1 000 generations.

\section{Results}

We found that for our neutral metacommunity model with realistic parameter values, regardless of its size, shape, and dispersal rate, the local species richness never exceeds a few species. Communities are dominated by one species, with sometime a few individuals from other species (Fig.1). For all values of $\omega$, regional species diversity at equilibrium is equal or below the number of communities. Unsurprisingly, we find that reducing the average migration rate increases the speciation rate, but it also increases the number of extinctions. For $\omega < 10^{-5}$, the regional species richness stabilizes at $n$, while for $\omega > 10^{-3}$, the entire metacommunity supports only a single species. We find a threshold migration rate around $\omega = 10^{-4}$ where the regional species richness increases suddenly. Around this value, the number of species varies between 1 and $n$. When $\omega < 10^{-5}$, the communities are so isolated that they are dominated by a single species, sometime with a small number of individuals from one or two other species.

We studied the effect of increasing the mutation rate beyond realistic values. Keeping $J_x$ at $10^4$ and $\omega = 5\times 10^{-4}$, we ran simulations for several mutation rates. Even a tenfold increase in the mutation rate ($\mu = 10^{-3}$) has little effect on the equilibrium regional species richness. The metacommunity sustains higher diversity around a mutation rate of $\mu = 10^{-2}$. This mutation rate is well above the typical mutation rate \cite{dra98,gav04}. This finding lends credit to the theory that the NTB requires unrealistically high speciation rates \cite{ric03}.

Within-population selection has an important impact on species richness (Fig. 2). We explored the parameter space to find the values of $\omega$ and $n$ (the number of communities) where diversity is highest. For $\omega$, diversity peaks around $5\times 10^{-4}$, with little variations between the different community shapes. Local diversity increases with $n$ betwen 1 and 5 but quickly reaches a plateau. Except for the complete metacommunity and the star, increasing $n$ beyond 10 has no effect on local diversity. Star, circle and random metacommunities share a similar regional species abundance distribution, which is lognormal-like with negative skewness (a long left tail). The complete metacommunity supports much less species especially as $n$ increases. There are fewer rare species than the regional distribution seen in the NTB, supporting the criticism of Ricklefs \cite{ric03}, which argued that the NTB predicted too many rare species. 

\begin{figure}[ht!]
	\centering \includegraphics[width=0.45\textwidth]{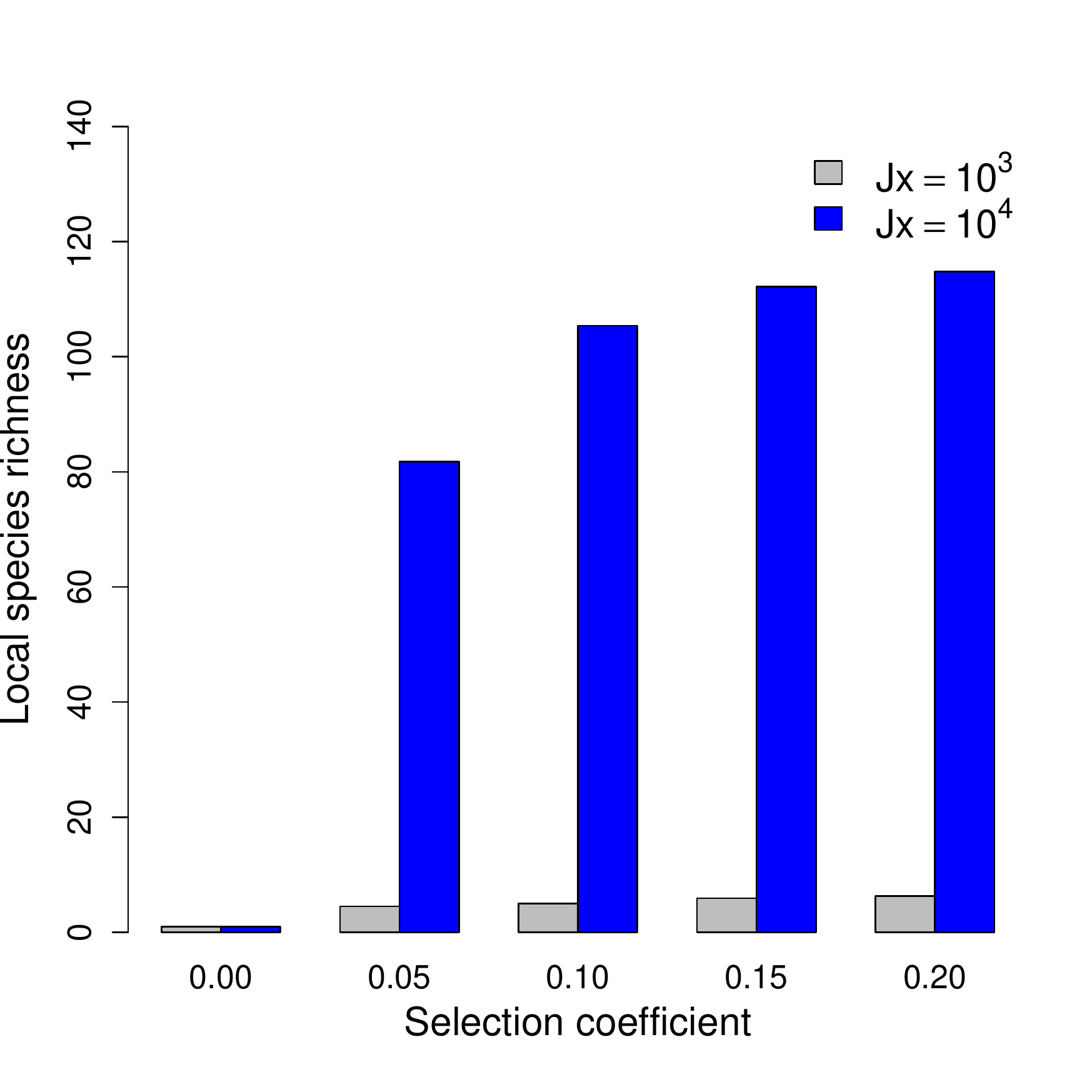}
	\caption{
The average number of species in local communities at equilibrium in the pseudo-selection model increases non-linearly with selection. The number of species quickly increases between $s = 0.00$ and $s = 0.05$, in part because selection pushes the alleles toward speciation but also because it reduces the fitness of migrants. We used random geometric graphs with $\omega = 5\times 10^{-4}$ and $n = 10$.
}
\end{figure}

We illustrated the effect of selection on species abundance distribution with a comparison to the tropical forest in Barro Colorado Island, Panama \cite{con02}. This plot of 50 ha contains 21 457 individuals and 225 species \cite{con02}. To find the minimal amount of selection required in our model to reach this level of diversity, we use the most speciose combination of parameters for random geometric graphs (Fig. 3) and then increase $s$ to reach an average local species richess of 220 using communities of size $J_x = 22 000$. This point is never reached. Local diversity increases with $s$ but saturates around $s = 0.15$ (Fig. 3). Selection above $s = 0.15$ has little effect on species richess but decreases the median lifespan from 380 with $s = 0.15$ to 240 with $s = 0.35$. The median population size at speciation decreases slightly with selection but remains in the 550-600 range for $0.35 > s > 0.15$. A small number of communities supported a significantly higher diversity than average. 5\% of the communities with $s = 0.25$ and $s = 0.35$ supported more than 200 species, with the most diverse local community supporting 233 species.

\begin{figure}[ht!]
  \centering
    \includegraphics[width=0.50\textwidth]{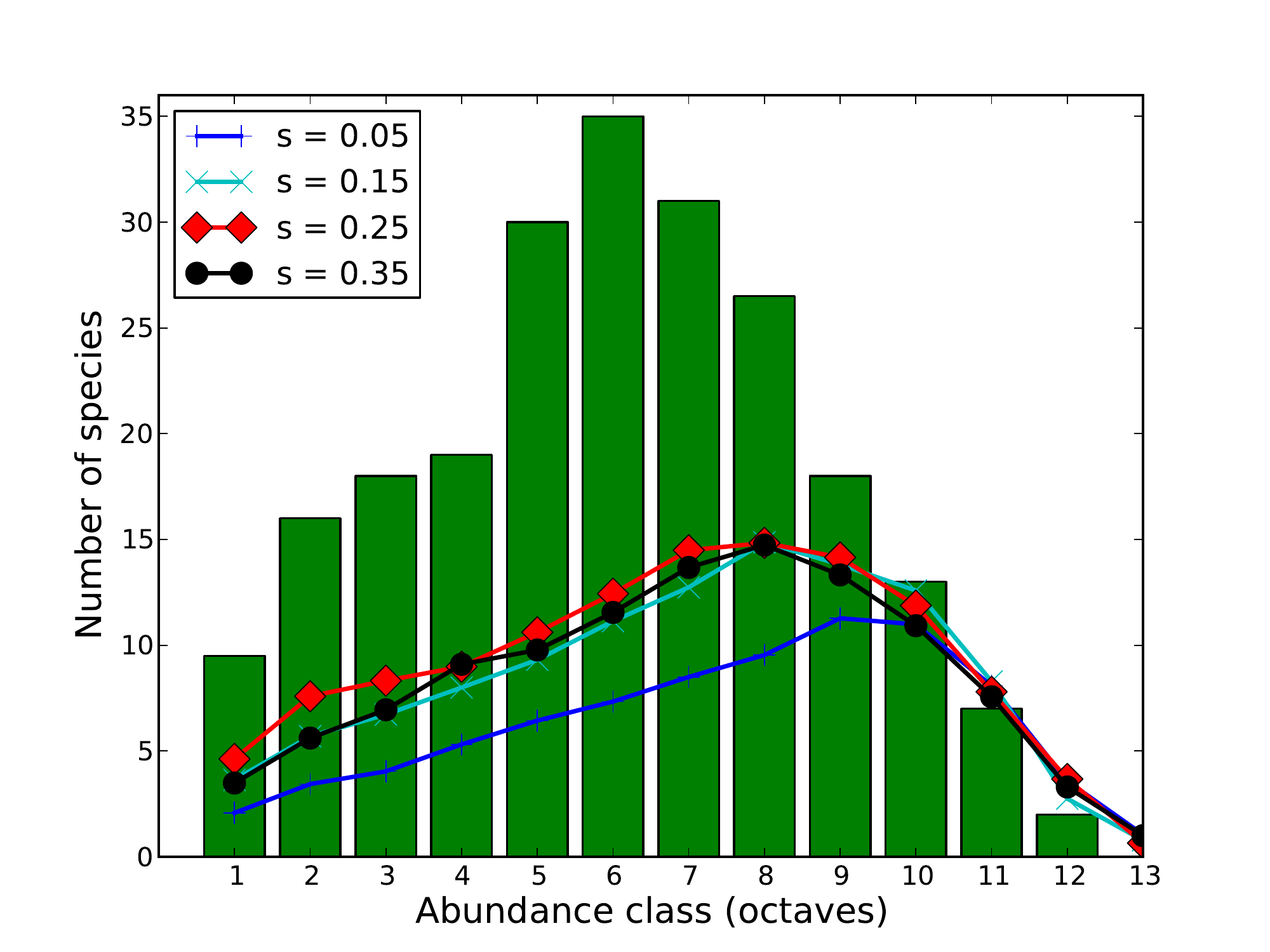}
	\caption{
Average local species abundance with selection from $0.05$ to $0.35$. Random geometric graphs were used with the most favourable set of parameters found ($\omega = 5\times 10^{-4}$ and $n = 10$). We set the local community size $J_x$ to 22 000 and the species distributions are compared to a 50 ha plot in Barro Colorado Island, Panama \cite{con02}. Selection increases local diversity but saturates quickly. Despite the favorable parameters and strong within-population selection, the level of diversity found in the Barro Colorado Island is only reached in less than 5\% of the communities subjected to strong selection ($s = 0.25$ and $s = 0.35$).
}
\end{figure}

\section{Discussion}

In this study, we developed a framework to study speciation as a population-process within neutral ecology. The speciation rate was not assumed to take any particular value, it is an emergent property of the system. It depends on selection, the mutation rate, migration, and the the shape of the metacommunity. Also, we made no assumption about the relationship between abundance and the speciation rate. Species with more individuals are likely to occupy more communities, so they will have more opportunities to speciate, but the relationship will depend on the shape of the metacommunity and the spatial distribution of the populations. Our goal was to examine the relationship between neutral ecology and a more mechanistic model of speciation based on population genetics. Phenomenological models are not inherently inferior \cite{mcg10} but they should be confronted to their mechanistic counterpart to determine if they can provide a good approximation of reality, under what conditions this approximation can hold, and what kind of assumptions are required to make it hold. Our assumptions deliberately made speciation easy to achieve. We used the BDM model with only two steps required to reach speciation and we ignored back mutations \cite{gav04}. The mutation rate chosen was plausible but high \cite{dra98,kum02}. There was always a mutation toward speciation available, arguably the most unrealistic assumption as the conditions for speciation are seldom common \cite{coy04}. All these assumptions greatly favor speciation, yet the model failed to produce metacommunities with many species unless selection is added or the mutation rate is set to impossible levels. The only element that could have a significant negative effect on speciation is the assumption that new migrants always carry the $a_xb_x$ haplotype \cite{gav04}, although this assumption is supported by empirical evidences against speciation with gene flow \cite{coy04}. The level of diversity seen in the BCI dataset is hard to reach with allopatric and parapatric speciation events within our neutral model, even with the addition of within-population selection. Yet, recent studies have improved the speciation model within neutral ecology \cite{ros10,ros11} and it remains to be seen if allopatric or parapatric speciation can be included explicitely in a neutral model without resulting in species-poor communities. This is an important challenge for the neutral theory given the importance of these modes of speciation. Adding temporal variations in $\omega$ might increase diversity to more reasonable levels as allopatric speciation events often follow changes in the environment \cite{coy04}.

Speciation can be achieved easily if mutations toward speciation are given some positive selection coefficient. But new species, being the result of the accumulation of fitness-enhancing mutations, should have greater fitness, which would violate the NTB's ecological equivalence assumption. Zhou and Zhang's nearly neutral model showed that small differences between species lead to markedly different species distributions \cite{zho08}. However, Du et al. \cite{du_11} argued that negative density dependence can offset the effect of competition and lead to neutral patterns. There is little doubt that selection plays an important role in speciation events \cite{coy04} and few neutral models of speciation have been developed \cite{nei83}. Speciation by drift alone is simply too slow \cite{tur01}. A possible alternative is to model speciation with positive assortative mating, in which individuals mate more often with similar individuals \cite{fel81,coy04}. This has been attempted in two recent models \cite{dea09,mel10}. De Aguiar et al. \cite{dea09} argue that biodiversity can arise without physical barriers. Although De Aguiar et al. \cite{dea09} use assortative mating in a grid to model spatial patterns of diversity, it is generally associated with sympatric speciation \citep[p.130]{coy04}. Melian et al. \cite{mel10} reached the conclusion that frequency-dependent selection lead to more species than neutral models.

When comparing models, one aspect to consider is their complexity. Theoretical populations genetics is mostly based on mathematical models that are simple enough to be analytically tractable, which has lead to a tendency to ignore spatial complexity \cite{epp10}. As allopatric and paratric speciation events rely on this spatial complexity, we have few theoretical models to study the effect of these forms of speciation on diversity. We chose to base our theory on the most common forms of speciation and introduced a simple method to model allopatric and parapatric speciation in complex spatial structures. While using graphs add a layer of complexity to neutral ecology, our approach fixes some of the problems of the point speciation model without adding new parameters for speciation, as we replace the speciation rate $v$ with the mutation rate $\mu$. More importantly, this approach allows us to divide a species in populations, a fundamental unit in evolution. Ricklefs \cite{ric03} argued that new species under the point speciation model would not be recognized as species, because those species have appeared instantaneously and are likely too similar. More importantly, one of the problems with a fixed speciation rate $v$ is that speciation is directly influenced by ecological factors such as isolation and habitats. In particular, the inhibiting effect of gene flow on speciation is ignored in most community models with speciation \cite{hub01,vol05,eti07,ros10} (but see \cite{ros11}), despite the fact that gene flow shapes patterns of speciation \cite{coy04}, which in turn has an important influence on the predictions \cite{eti07,eti11}.

\section{Acknowledgements}

We thank James Rosindell for many stimulating discussions and helpful comments. We also thank Egbert Giles Leigh Jr., Rampal Etienne, Evan Economo, Nicolas Loeuille, and Franck Jabot for reviewing an early draft of this article. We thank the volunteers who spent days collecting the data on Barro Colorado Island, Panama. This work was supported by a research grant from the Canada Research Chair program to DG.

\bibliographystyle{plain}
\bibliography{phdp}

\end{document}